\documentclass[aps,showpacs,twocolumn]{revtex4} 
\usepackage{amssymb}
\usepackage{amsmath} 
\usepackage[dvips]{graphicx}
\usepackage{multirow}
\usepackage{natbib}
\bibpunct{}{}{,}{n}{}{=}

\newcommand{\Tr}{\mathrm{Tr}}

\newlength{\barheight}

\newenvironment{packed_enum}{
\begin{enumerate}
  \setlength{\itemsep}{0.5pt}
  \setlength{\parskip}{0pt}
  \setlength{\parsep}{0pt}
}{\end{enumerate}}

\newcommand{\oldcite}[1]{[\cite{#1}]}
\newcommand{\mycite}[2]{[\cite{#1}#2]}

\begin{document}
 
\title{Error characterization in Quantum Information Processing: a protocol for analyzing spatial correlations and its experimental implementation}
\author{Cecilia C. L\'opez, Benjamin L\'evi, David G. Cory} 
\affiliation{Department of Nuclear Science and Engineering, MIT, Cambridge, MA 02139, USA} 

\begin{abstract}
We present a protocol for error characterization and its experimental implementation with 4 qubits in liquid state NMR.
The method is designed to retrieve information about spatial correlations and 
scales as $O(n^w)$, where $w$ is the maximum number of qubits that have non-negligible interaction. 
We discuss the practical aspects regarding accuracy and implementation.
\end{abstract}


\maketitle

\section{Introduction}

Precise and reliable control of the system constituting a quantum information processor (QIP) remains
one of the biggest challenges in the quantum information field. 
In order to assess the reliability of a device and to tailor quantum error correction schemes to a faulty one, we need to characterize the errors occurring in the system. 
Quantum Process Tomography (QPT) \oldcite{chuang} presented a first answer to this problem, providing full characterization of the process under analysis. 
Experimental implementations of QPT have been already conducted in a variety of small systems \oldcite{qft,blatt,steinberg,solidstate}.  
Nevertheless, QPT becomes impractical beyond a few qubits, as it requires $O(2^{4n})$ experiments for a system of $n$ qubits.
In recent years, the idea of getting less information at a lower cost has become a popular strategy to tackle error characterization,
and several works have been devoted to the subject \oldcite{emersonfirst,ourpaper,emerson,knill,paz,lidar,njp}. Our proposal fits in this context, 
providing a subset of information yet still using scalable resources.

The scheme we present selectively keeps information 
about the spatial correlations of the errors occurring in the process under study (a gate, a noisy channel, etc.). 
Both the magnitude and the structure of the errors are relevant to evaluate fault-tolerance.
In particular, fault-tolerance
threshold theorems are designed for certain conditions of spatial correlation (also termed range or locality) \oldcite{locality}.
So even once it is experimentally determined that only up to $w$ qubits are involved in an error process, 
we have to further establish in which way the
${n \choose w}$ possible sets are being affected.

We have implemented our protocol in a liquid state NMR 4-qubit QIP. 
The core mathematical work for this protocol was introduced in \oldcite{ourpaper}. Here we extend our 
proposal to a more general setting and include an experimental realization.
Our basic method, like others proposed 
\oldcite{chuang,emersonfirst,ourpaper,emerson,paz,lidar}, assumed error-free implementation stages.
This idea is of course unrealistic in practice, and implementation errors complicate the task of 
reliable error characterization. Thus here we have included an analysis of their effect.

\section{Theory behind the protocol} 
We start by describing the action of a general map on the state
of an $n$-qubit system ($D=2^n$), described by an initial state $\rho_0$, as
\begin{eqnarray}
\mathbb{S}(\rho_0) = \int p(\vec{\eta}) \ E(\vec{\eta}) \ \rho_0 \ E^{\dagger}(\vec{\eta}) \ d\vec{\eta}.
\label{errormap}
\end{eqnarray}
The vector $\vec{\eta}$ denotes $D^2$ complex coefficients $\{\eta_0, \eta_l, l=1,\ldots,D^2-1\}$ that 
parametrize $E$, 
an arbitrary operator in the Hilbert space $\mathcal{H}_D$, as
\begin{eqnarray}
E = \eta_0 I + \sum_{l=1}^{D^2-1} \eta_l \ O_l, \ \ O_l = \bigotimes_{j=1}^n O_l^{(j)} \label{expansionE}
\end{eqnarray}
where each $O_l^{(j)}$ is an element of the Pauli group $\{I, \sigma_x, \sigma_y, \sigma_z\}$, but at least one factor in each $O_l$ is a Pauli matrix. 
Notice that $\Tr[O_l O_{l'}]=D \delta_{l,l'}$. 
When $p(\vec{\eta})$ is a (real) nonnegative distribution, eq. (\ref{errormap}) 
describes an arbitrary 
completely positive (CP) map, and the condition $\int p(\vec{\eta}) |\vec{\eta}|^2 d\vec{\eta} = 1$ guarantees the preservation of $\Tr[\rho]$.
This representation of a CP map is an operator 
sum representation with continuous parameters. We prefer this form as it's more suitable to describe non-unitary dynamics arising 
from stochastic Hamiltonians. Our parameters, the $\vec{\eta}$, are trivially related to the parametrizations of the operator sum representations
used in other works \oldcite{emerson,paz}.
Also, for small $\eta_l$, it's possible to relate (\ref{errormap}) to a 
description of the noise in terms of generators rather than propagators \oldcite{ourpaper}.

When necessary, we shall denote the $\eta_l$ in more detail as $\eta_{j,k...}^{p,q...}$, where $j > k > \ldots$ label qubits, and $p,q,\ldots={x,y,z}$. 
Therefore $\eta_{j,k...}^{p,q...}$ labels a term in eq. (\ref{expansionE}) that is a product of $\sigma_p$ 
for qubit $j$, $\sigma_q$ for qubit $k$, etc., and $I$ for the qubits absent in the subscript.
Notice that the number of qubits in the subscript gives the Pauli weight (also Hamming weight) of the term \oldcite{pauliweight}.\\

In our protocol we measure a subset of $m$ qubits at a time, which will allow us to extract the magnitude of the errors involving that subset.
So we now break the system in two: $m$ qubits belonging to the Hilbert space 
$\mathcal{H}_M$, which will be the ones measured, 
and $\overline{m}=n-m$ qubits belonging to the complementary space $\mathcal{H}_{\overline{M}}$, 
$\mathcal{H}_D = \mathcal{H}_M \otimes \mathcal{H}_{\overline{M}}$. 
We require the initial state to be separable in these two spaces, and within $\mathcal{H}_M$, 
thus 
\begin{eqnarray}
\rho_0 = \rho_0^{(M)} \otimes \rho_0^{(\overline{M})} \ \ \textrm{and} \ \
\rho_0^{(M)} = \bigotimes_{j \in M} \rho_0^{(j)}. 
\end{eqnarray}
We now perform a $U(2)^{\otimes m}$ twirl on the target map $\mathbb{S}$,
\begin{eqnarray}
\rho_1= \frac{1}{K^m} \sum_{k = 1}^{K^m} \mathcal{C}^\dagger_k \mathbb{S}\left( \mathcal{C}_k \rho_0 \mathcal{C}_k^\dagger \right) \mathcal{C}_k
\label{twirl}
\end{eqnarray}
where the $\mathcal{C}_k$ are $m$-fold tensor products of the twirl operators on one qubit, for the $m$ qubits in $\mathcal{H}_M$.
For the task we propose, a subset of $K=6$ operators from the Clifford group will suffice.
We refer to this minimum set of operators required to apply the twirl (\ref{twirl}) as the $6^m$-Clifford element pool (see Appendix A for references and some mathematical details on this twirling).

The reduced density matrix $\rho_1^{(M)}=\Tr_{\overline{M}}[\rho_1]$ is the state of the $m$ qubits we measure.
In general, the fidelity decay, which gives a measure of how 
$\rho_1$ differs from the initial $\rho_0$, encodes information about the map $\mathbb{S}$ we are characterizing \oldcite{emersonfirst,ourpaper}. In particular, considering only the qubits we will measure and expressing 
$\rho_0^{(\overline{M})}=(I^{\otimes \overline{m}}+\rho^{(\overline{M})}_{dev})/2^{\overline{m}}$, we obtain
\begin{eqnarray}
\Tr[(\rho_0^{(M)})^2] - \Tr_M[ \rho_0^{(M)} \rho_1^{(M)}] = \gamma^{(M)} - \Omega_{dev}^{(M)} \label{fiddecay} \\
\gamma^{(M)} = \sum_{l} \langle | \eta_l |^2 \rangle 
\left(\prod_{j \in M} \mathcal{P}_j - \prod_{j \in M} \mathcal{C}_j(l)  \right) \label{gamma}
\end{eqnarray}
and $\Omega_{dev}^{(M)}=0$ if $\rho^{(\overline{M})}_{dev} =0$ 
($\Omega_{dev}^{(M)}$ for $\rho^{(\overline{M})}_{dev} \neq 0$ can be exactly computed if desired). 
The quantity defined in the LHS of 
(\ref{fiddecay}) is the fidelity decay rate \oldcite{emersonfirst,ourpaper} we will analyze in this work.
In eq. (\ref{gamma}) we have denoted $\langle \ldots \rangle = \int p(\vec{\eta}) \ldots d\vec{\eta}$,
$\mathcal{P}_j=\Tr[(\rho_0^{(j)})^2]$ the initial purity of each of the $M$-qubits, and
\begin{eqnarray}
\mathcal{C}_j(l) = \left\{ \begin{array}{ll} 
          (2/3) (1 - \mathcal{P}_j/2) & \textrm{ if $O_l^{(j)}= \sigma_x, \sigma_y, \sigma_z$}\\
          \mathcal{P}_j & \textrm{ if $O_l^{(j)}= I$} \end{array} \right.
\end{eqnarray}
For the derivation of eqs. (\ref{fiddecay}-\ref{gamma}) we use the equivalence between a Clifford twirl and a Haar twirl \oldcite{dankert}, and 
apply the tools developed in \oldcite{benstrick}. See also Appendix A.\\

In the sum (\ref{gamma}), a $l$-term vanishes when $O_l$ is the identity operator for the $m$ qubits being measured. 
So by choosing different sets of  $M$-qubits, it is possible to leave out certain $\eta_l$ in a given $\gamma$.
Note however that $\mathcal{C}_j$ does not distinguish 
the direction of the Pauli matrices, so there is an implicit coarse-graining of all the $O_l$ that have the identity $I$ 
for the same subset of $m$ qubits. This is how we get the ``collective 
coefficients'' $\eta^{col}_l$,
\begin{eqnarray}
|\eta^{col}_j|^2 = \sum_{p=x,y,z} |\eta_j^p|^2, \ \ |\eta^{col}_{j,k}|^2 = \sum_{p,q=x,y,z} |\eta_{j,k}^{p,q}|^2, \ \textrm{etc.} 
\label{collective}
\end{eqnarray}
For example, the terms $\sigma_x^{(1)}\sigma_x^{(3)}$ and $\sigma_x^{(1)}\sigma_z^{(3)}$ contribute collectively to $\eta^{col}_{1,3}$.

By combining the $\gamma$'s from different $M$-sets it is possible to further isolate the collective coefficients. 
If we prepare $a$ and $b$ in a pure state, so $\mathcal{P}_a = \mathcal{P}_b=1$, we obtain
\begin{equation}
\frac{9}{4}\!\left(\!\gamma^{(a)} \!+\! \gamma^{(b)} \! - \! \gamma^{(a,b)}\!\right)\! 
= \! \langle|\eta^{col}_{a,b}|^2\rangle + \sum_j \langle|\eta^{col}_{a,b,j}|^2\rangle + ... \!
\label{eta12}
\end{equation}
(Please refer to Appendix B for the formulae of the $\gamma$'s involved in this example).
If 3-body and higher multi-body terms can be neglected, eq. (\ref{eta12}) gives $\langle|\eta^{col}_{a,b}|^2\rangle$.
Similarly, the combination of the seven $\gamma^{(j)}$, $\gamma^{(j,j')}$ and $\gamma^{(j,j',j'')}$ for a set of three qubits $a,b,c$ would return 
$\langle|\eta^{col}_{a,b,c}|^2\rangle$, and so on. The collective coefficients then report on the spatial correlations of errors.

\section{The protocol} 

We present now a systematic protocol for measuring the collective coefficients involving the $m$ qubits of a particular subset.
\begin{packed_enum}
\item[$i.$] Prepare each of the qubits to be measured in the initial state $|0\rangle$.
Prepare each of the other $\overline{m}$ qubits in the maximally mixed state $I/2$. 
\item[$ii.$] Apply one of the $m$-fold Clifford operators from the $6^m$-Clifford element pool.
\item[$iii.$] Implement the target gate or noise under study.
\item[$iv.$] Invert the Clifford operator applied in $ii$.
\item[$v.$] Measure the projection of the resulting state on the initial state $|0\rangle$, for each of the qubits being measured. 
\end{packed_enum}
To implement the twirl, we repeat $i-v$ each time taking a different operator from the $6^m$-Clifford element pool, and average the results.

In a canonical QIP, the implementation of this protocol to measure the decay rates involving $m$ qubits will
require $N$ realizations. This will take care of: $1)$ preparing the desired initial state as in step $i$ 
(starting from the $|0\rangle^{\otimes n}$ state, and randomly 
flipping the $\overline{m}$ qubits we don't measure); $2)$ measuring the fidelity decay through repeated projective measurements of the $m$ qubits 
as prescripted in step $v$ (notice eq. (\ref{fiddecay}) becomes $1 - \Tr_M[ \rho_0^{(M)} \rho_1^{(M)}] = \gamma^{(M)}$ for the proposed initial state); 3) implementing the twirl approximately, by randomly drawing the twirl operators for steps $ii$ and $iv$ from the corresponding $6^m$-Clifford element pool (or from an infinite pool of one-qubit random rotations).

With this strategy, the outcome of the measurement step $v$ is a Bernoulli variable, and thus $N$ can be estimated from usual statistics. More precisely, the standard error in our estimation of the decay rate will be $\sigma_\gamma \leq 1/\sqrt{N}$ (following the Central Limit Theorem), so for a desired precision $\delta$ we must have $N \geq \delta^{-2}$. On the other hand, 
the Chernoff Bound tells us that 
for a desired precision $\delta$ and an error probability $\epsilon_N \ll 1$,
we must have $N=\log(2/\epsilon_N)/(2\delta^2)$, which is a stronger requirement when $\delta < 2 e^{-2}$. 
In any case, $N$ is independent of the number of qubits thus the efficiency of the protocol is independent of the size of the system.

In the case of liquid state NMR ensemble QIP, 
pseudo-pure state preparation allows for initialization in the $I/2$ state over the ensemble of molecules, 
and ensemble measurements avoid the need 
of repeated realizations in order to perform step $v$ by Quantum State Tomography. This is the case in our experiment.\\

The measurement step $v$ retrieves the information to calculate the decay rates for the $m$ qubits and for any smaller 
subset of qubits within those. For example, twirling qubits $a$ and $b$ only ($m=2$), we can obtain 
$\gamma^{(a)}$, $\gamma^{(b)}$ and $\gamma^{(a,b)}$
in one shot, and calculate $\langle |\eta_{a,b}^{col}|^2 \rangle$ as in eq. (\ref{eta12}) neglecting the higher multi-body terms. 
This procedure can be repeated for the ${n \choose 2} = n(n-1)/2$ pairs of
qubits, and by doing so all the collective coefficients for 1- and 2-body terms can be extracted.

The scalability of the method goes as follows. 
If we can neglect the multi-body terms above a certain Pauli weight $w$, and $N$ is the number of realizations required to measure the 
fidelity $\gamma$ for $w$ qubits, then with $N \ {n \choose w} \leq N n^w/w!$ experiments
we can estimate all the non-negligible coefficients. This should be compared against $N 2^{4n}$, the overhead in the number of experiments required 
for QPT. 
We emphasize here that our proposal seeks to characterize the correlations among up to $w$ qubits 
in order to establish the range of the noise - not to characterize the process fully.

To use our protocol, the negligibility of multi-body terms above a certain Pauli weight $w$ must be established a priori.
In a canonical QIP we could apply our protocol to measure all the $n$ qubits, obtain all the decay rates 
$\gamma^{(j)},\ \gamma^{(j,k)}, \ldots, \gamma^{(1,\ldots,n)}$ and extract all the collective coefficients after only $N$ experiments 
(independently of $n$). In this way we can handle all the Pauli weights, from $0$ to $n$.
But the error in the decay rates $\sigma_{\gamma}$ propagates into the $\eta^{col}$ for $m$ qubits inefficiently, roughly as 
$\sigma_{\eta^2} \propto \sqrt{\sum_{j=0}^w {n \choose j}} \sigma_{\gamma}$.
Therefore our strategy of looking for spatial correlations after establishing a cut-off Pauli weight $w$:
to our knowledge, neither QPT nor other proposals so far
are able to make use of the negligibility of high order correlations in order to gain further insight.

An approach to establishing this cut-off $w$, demanding the same resources as our protocol, is to apply the method developed 
by Emerson \textit{et al.} \oldcite{emerson}
which gives the probability of errors happening distinguishing them only by Pauli weight 
(this is, an average of all the ${n \choose w}$ collective coefficients having Pauli weight $w$). 
It is worth pointing out that both methods
require the same experimental work: the algorithms retrieve different information because the measurement and processing of the data
is different, but actually both protocols can be used complementarily.

\section{Experimental results} 

We implemented our protocol in a liquid state NMR QIP using the four $^{13}C$-labeled carbons of 
crotonic acid \oldcite{crotonic} in a 400MHz Bruker spectrometer.
The initial state preparation and all the gates required by the protocol were implemented with RF pulse sequences engineered using either 
GRAPE \oldcite{grape} or SMP \mycite{crotonic}{(c)} methods. Their simulated gate fidelities $F_g$ were on average 0.98. 
The typical experimental performance of one-qubit gates on the
spectrometer is 1-2\% below their simulated fidelities.
See \oldcite{qft} for details on the model used in the simulation.\\

We studied the following processes (see Table \ref{table:results} for more details): 
$i)$ A time suspension sequence $I_E$, since it is important 
to study our ability to ``do nothing'' in a system with a natural Hamiltonian that is always on.
$ii)$ An engineered error creating a coupling between qubits $1$ and $2$, of the form $C_{12}(\beta)=\exp(-i \beta \sigma_z^{1} \sigma_z^{2})$.
We chose $\beta=0.1$, and $\beta=0.4$ (the previous one applied consecutively 4 times).
$iii)$ A $CNOT$ gate between qubits $1$ and $2$: 
$CNOT = 0.5 (I+\sigma_z^{(1)}+\sigma_x^{(2)} - \sigma_z^{(1)}\sigma_x^{(2)}$).
Also, this gate applied twice: $CNOT^2=I$.
These gates are more complex than one-qubit operations (which are typically less than $1$ msec long)
and they all involve refocusing idle times (periods of free evolution under the internal Hamiltonian) in their pulse sequences. 

The results on the measurement of the collective coefficients for the qubit pairs $(1,2)$, $(2,3)$ and $(1,4)$ are presented in Table \ref{table:results}.
The pair $(1,2)$ is the one targeted by the $C_{12}$ and $CNOT$ gates, while the other two are the pairs involving qubits $1$ or $2$ 
with the highest $J$-coupling \oldcite{crotonic}.
We expect the errors for the three chosen pairs 
to be larger (due to internal evolution that is not perfectly refocused).
See Appendix C for more details on the experiment and simulations using liquid state NMR QIP.\\

\begin{table}[h]
\begin{center}
\begin{tabular}{c|c|c|c|c}
\hline\hline
\multicolumn{2}{c|}{Gate} & \hspace{0.2cm} $\langle|\eta_{1,2}^{col}|^2\rangle$ \hspace{0.2cm} & 
\hspace{0.2cm} $\langle|\eta_{2,3}^{col}|^2\rangle$ \hspace{0.2cm} &
\hspace{0.2cm} $\langle |\eta_{1,4}^{col}|^2\rangle$ \hspace{0.2cm} \\
\hline\hline
\multirow{4}{1.7cm}{$\ I_E \ \ \ \ \ \ $ $12.2$ msec $\ \ F_g=0.96$} & $meas$ & 0.02 & 0.02 & 0.01 \\
 & $theo$ & 0.00 & 0.00 & 0.00\\
 & $sim G$ & 0.01 & 0.02 & 0.00\\
 & $sim E$ & 0.01 & 0.03 & 0.00\\
\hline
\multirow{4}{1.7cm}{$\ C_{12}(0.1)$ $4.88$ msec $\ \ F_g=0.99$} & $meas$ & 0.02 & 0.02 & 0.01 \\
 & $theo$ & 0.01 & 0.00 & 0.00\\
 & $sim G$ & 0.02 & 0.00 & 0.00\\
 & $sim E$ & 0.02 & 0.00 & 0.00\\
\hline
\multirow{4}{1.7cm}{$\ C_{12}(0.4)$ $19.52$ msec $\ \ F_g=0.87$} & $meas$ & 0.26 & 0.03 & 0.03 \\
 & $theo$ & 0.15 & 0.00 & 0.00\\
 & $sim G$ & 0.23 & 0.01 & 0.00\\
 & $sim E$ & 0.24 & 0.02 & 0.02\\
\hline
\multirow{4}{1.7cm}{$\ CNOT$ $11.88$ msec $\ \ F_g=0.99$} & $meas$ & 0.32 & 0.01 & 0.03 \\
 & $theo$ & 0.25 & 0.00 & 0.00\\
 & $sim G$ & 0.25 & 0.01 & 0.00\\
 & $sim E$ & 0.28 & 0.02 & 0.00\\
\hline
\multirow{4}{1.7cm}{$\ CNOT^2$ $23.76$ msec $\ \ F_g=0.97$ } & $meas$ & 0.07 & 0.05 & 0.04 \\
 & $theo$ & 0.00 & 0.00 & 0.00\\
 & $sim G$ & 0.01 & 0.01 & 0.00\\
 & $sim E$ & 0.01 & 0.02 & 0.00\\
\hline
\end{tabular}
\caption{Measured ($meas$) collective coefficients for selected pairs of qubits, for the various gates studied.
$theo$ are the theoretical values for these gates as described in the text.
The $sim$ values come from numerical simulation: $sim G$ are the values obtained by the calculation of the propagator
as given from the RF pulse sequence corresponding to the gate ($G$) alone; $sim E$ are the values obtained by completely mimicking the experiment ($E$), 
considering the average of 36 simulations, each applying RF pulses to prepare the initial state and implement the Cliffords gates 
and the gate under study.}
\label{table:results}
\end{center}
\end{table}

The resulting $\langle|\eta_{a,b}^{col}|^2\rangle$ for the various pairs of qubits shown in Table \ref{table:results} exhibit good agreement with the predicted ones. 
Notice that the largest differences between measured and predicted appear on the pair $(1,2)$ 
and on the most complex gates: $CNOT^2$, $CNOT$, $C_{12}(0.4)$. 
This indicates that these deviations are due to the errors 
expected from spurious processes in our QIP, particularly an imperfect refocusing during the gate sequence, 
rather than from an imperfect implementation of the protocol.

The simulations account for well-known sources of error in liquid NMR QIP (imperfect pulse design \mycite{crotonic}{(c)}, RF field inhomogeneities \oldcite{qft}, and the presence of the magnetically active hydrogens in Crotonic Acid \oldcite{crotonic}). It worth mentioning that the main errors ocurring in one-qubit gates are correlated one-body errors (this is, a one-qubit rotation that is slighly off), and they do not introduce two-body errors, which are the main target of a spatial correlation analysis. This can be noticed also 
in the fact that the $sim G$ values are similar to the $sim E$ values (a change of $0.00-0.01$, except in two cases where we found $0.02$ and $0.03$).
Moreover, there is a contribution arising from $T_2$ relaxation. However, a calculation over the theoretical propagator 
and the numerics over the simulated one show a change on the order of $0.01$.

Elements outside our model of the system, which would explain futher the gap between theory and experiment, are $B_0$ (static) field inhomogeneities,
the presence of transients and residual non-linearities in the spectrometer circuitry, and an imperfect spectral fitting of the measured signal. These
are well-know issues in liquid NMR QIP, whose effect falls within the $1-2\%$ error.\\

We must differentiate between the implementation errors in the protocol (initial state preparation, one-qubit twirling and readout), 
and the errors in the gate under study.
The former ones affect the accuracy of protocol, which accounts for the measured non-null coefficients that are expected to be zero. As discussed,
RF field inhomogeneities, the presence of hydrogens and $T_2$ relaxation already give an error bar $\sigma_{\eta^2} \approx 0.03$. And there are still 
other sources of error mentioned above that could make $\sigma_{\eta^2}$ larger, but within that order.

Moreover, a fiducial initial state preparation is critical to the sucess of the algorithm. We can quantify
this as follows. If we call $\varepsilon_0$ the error in the initial state preparation, and similarly we call $\varepsilon_1$ the error in the implementation of the 
Clifford gates, an error propagation in the formula for the $\gamma$'s gives 
$\sigma_\gamma^2 \leq \varepsilon_0^2 (1 + 4 \gamma)+\varepsilon_1^2$, and then it's simple to propagate this into the formulas
for the $|\eta^2_l|$: for example, for a pair of qubits $a,b$ we follow eq. (\ref{eta12}) and obtain 
$\sigma_{\eta^2} = \frac{9}{4} \sqrt{\sigma_{\gamma_a}^2 +\sigma_{\gamma_b}^2+ \sigma_{\gamma_{a,b}}^2}$. These $\varepsilon$'s
account for non-statistical errors (typically correlated one-body errors) and set the accuracy of the method.
Given the low complexity
of initial state preparation and one-qubit operators, these $\varepsilon$'s are smaller 
than the errors in target operations (a fact reflected, for example, in the gate fidelities). This is why even though the theory 
was developed for error-free initial state preparation and twirl, the actual implementation can still
retrieve information about the target operations.\\

\section{Discussion on this and other proposals}

Our protocol belongs to the family of characterization strategies 
based on the use of twirling to coarse-grain the original $O(2^{4n})$ complex parameters to a $poly(n)$ 
number (\textit{cf} the work by Emerson \textit{et al.} in \oldcite{emersonfirst,emerson}). In our case, we gain detail about 
the process under study (spatial correlations).
Note also that 
the required resources in our proposal are within the minimal performance 
expected from a functional QIP: fiducial state preparation and readout, and a set of $6n$ one-qubit gates.

One of the protocols presented by Bendersky \textit{et al.} in \oldcite{paz} 
also returns similar information (although our method performs a coarse-graining of the 
directions). Nevertheless,
their proposal requires more demanding resources 
(although still scaling as $poly(n)$): a full twirl on $U(D)$ and the implementation of the $O_l$ operators.

Other characterization methods include the ancilla-assisted ones in \oldcite{paz,lidar}, but unfortunately they have a rather strong requirement: 
one or more clean error-free qubits within the system.
Contrasting with all the proposals discussed so far,
the method developed by Knill \textit{et al.} in \oldcite{knill} does not
require error-free stages, allowing for certain type of errors to occur during the whole computation.
Unfortunately it's not yet evident how to take their scheme beyond one-qubit QIP \oldcite{njp}.\\

On a different note, we would like to point out a particular feature of the variables $\langle |\eta_l|^2 \rangle$, which are 
the diagonal elements of the so-called $\chi$-matrix in the $O_l$ basis (\textit{cf} the $a_i$ in \oldcite{emerson}, 
or the $\chi_{m,m}$ in \oldcite{paz}). 
When the error $E$ acts on a short time
we can expect the coefficients to be small and eq. (\ref{expansionE}) can be taken instead as a first order 
Taylor expansion of $E$, where the $\eta_l$ with $l \geq 1$
play the role of a generator of the error. This idea was originally developed in \oldcite{ourpaper}
under the same setting, but aiming to study the generators directly led to limitations in the error model.
Nevertheless, drawing the connection between the two opens the possibility 
of identifying a generator, which is essentially a Hamiltonian with varying parameters $\eta_l$ whose
dynamics we can only observe on average through the $\langle |\eta_l|^2 \rangle$. 
This interpretation allows a different insight into 
the dynamics of the system, as many physical processes are better described by the action of a stochastic Hamiltonian rather
than by an operator-sum representation arising from a system+bath picture.\\

In conclusion, we have presented a method to characterize the spatial correlations occurring in a gate or
process under study, showing its potential through a liquid state NMR QIP experiment. 
Second, we have pointed out the need of experimental feedback in order to arrive to a not only scalable but feasible
protocol, as the one we introduced. Finally, we have analyzed the relevance of implementation errors, showing the need for strategies 
that are not only scalable but also robust.\\

We thank J. Emerson, M. Silva and L. Viola for useful discussions, and
J.S. Hodges and T. Borneman for invaluable assistance with the experiments.
This work was supported in part by the National Security Agency (NSA) under Army Research Office (ARO) contract number W911NF-05-1-0469.

\section{Appendix A: On the Clifford twirl}

Consider the twirling of the map $\mathbb{S}$ for $m$ qubits in the Hilbert space $\mathcal{H}_M$, of dimension $2^m$. 
As shown in \oldcite{dankert}, a Haar twirl is equivalent to Clifford twirl as follows
\begin{eqnarray}
\rho_1 &=& \int_{U(M)} dU U^\dagger \mathbb{S}\left( U \rho_0 U^\dagger \right) U  \label{twirl1} \\
&=& \frac{1}{K^m} \sum_{k = 1}^{K^m} \mathcal{C}^\dagger_k \mathbb{S}\left( \mathcal{C}_k \rho_0 \mathcal{C}_k^\dagger \right) \mathcal{C}_k
\label{twirl2}
\end{eqnarray}
where $dU$ denotes the Haar measure on the unitary group $U(2^m)$. The sum on eq. (\ref{twirl2}) runs over all the elements 
of the Clifford group for $U(2^m)$.
In particular, this equation holds for a twirl on $U(2)^{\otimes m}$,
where the $\mathcal{C}_k$ are $m$-fold tensor products of the $K=24$ elements of the Clifford group for one qubit, 
for the $m$ qubits in $\mathcal{H}_M$.\\

We can write the Clifford operators for one qubit as products $SP$ of an element $S$ of the Symplectic group 
$\{\exp(-i\nu (\pi/3)(\sigma_x+\sigma_y+\sigma_z)/\sqrt{3}),\ \nu = 0,1,2; \ \exp(-i(\pi/4)\sigma_p), \ p=x,y,z\}$
and an element $P$ of the Pauli group $\{I, \sigma_x, \sigma_y, \sigma_z\}$.
Due to some redundancy in the symplectic twirling, the minimum number of elements to achieve the 2-design property (eq. (\ref{twirl}), as defined in \oldcite{dankert})
is actually half of the whole Clifford group:
$S$ can be taken from either $S_1=\{\exp(-i \nu (\pi/3)(\sigma_x+\sigma_y+\sigma_z)/\sqrt{3}), \ \nu=0,1,2\}$
or $S_2=\{\exp(-i(\pi/4)\sigma_p),\  p=x,y,z\}$.

This number can be further decreased to $K=6$ if we take an initial state $\rho_0$ as described in the main text (eq. (3)) and
we remark that we will be only interested in the projection of $\rho_1^{(M)}$ onto 
given initial states of the form $\rho_0^{(j)} = (I+\sigma_z)/2$.
There is then an additional redundancy in the values of these projections which enables us to chose $P$
in the set $\{P_1,P_2\}$ with $P_1 = I$ or $\sigma_z$ and $P_2 = \sigma_x$ or $\sigma_y$. 
Here we have chosen $\rho_0^{(j)}$ along $z$ for definiteness (which is what we use in the experiment), 
but an equivalent result can be obtained for the $x$ or $y$ directions. 
Notice that the number of elements $K$ required in eq. (\ref{twirl2}) depends on what quantity we will measure.
In $U(2)^{\otimes m}$ (as it is our case), whether some elements are redundant can be proved simply by straightforward calculation. \\ 

Implementing a Haar twirl as defined on eq. (\ref{twirl1}) would require sampling over an infinite pool of random rotations $U_k$.
The equivalence with the Clifford twirl allows us to implement the twirling with a finite set of gates: the $K^m$ Clifford operators in $U(D)$. Nevertheless, although not infinite, the pool is of exponential size in $m$, thus again calling for a 
sampling strategy to implement the twirl approximately. Of course, for $m$ small enough, it may be possible to implement the twirl exactly 
(as in our experiment, where $m=2$ and the Clifford pool has size $36$). But on general grounds we can take on any of the two approaches. 

Depending on the experimental setup in question, one twirl may be more robust than the other (depending on what type of errors are expected 
for one-qubit gates), and also one pool may be easier to construct than the other.

\section{Appendix B: Combining $\gamma^{(a)}$, $\gamma^{(b)}$ and $\gamma^{(a,b)}$ to retrieve $\langle|\eta^{col}_{a,b}|^2\rangle$}

To illustrate the mechanism of combining different $\gamma$'s, consider measuring one and two qubits as follows. 
With $\mathcal{A}_j=(4\mathcal{P}_j-2)/3$:
\begin{widetext}
\begin{eqnarray}
\gamma^{(a)} = \mathcal{A}_a \bigg(
\langle|\eta^{col}_a|^2\rangle + \sum_{j \neq a} \langle|\eta^{col}_{a,j}|^2\rangle + 
\sum_{k > j \neq a} \langle|\eta^{col}_{a,j,k}|\rangle^2 + \ldots \bigg) \ \ \textrm{and similarly} \ \gamma^{(b)}. 
\nonumber \ \ \ \ \ \ \ \ \ \ \ \ \ \ \ \ \ \ \ \ \ \ \\
\gamma^{(a,b)} = 
\frac{8 \mathcal{P}_a\mathcal{P}_b + 2(\mathcal{P}_a +\mathcal{P}_b)-4}{9} \ \times \
\bigg( \langle|\eta^{col}_{a,b}|^2\rangle + \sum_{j \neq a,b} \langle|\eta^{col}_{a,b,j}|^2\rangle + \ldots \bigg) +
\nonumber  \ \ \ \ \ \ \ \ \ \ \ \ \ \ \ \ \ \ \ \ \ \ \ \ \ \ \ \ \ \ \\
\mathcal{A}_a \mathcal{P}_b \bigg( \langle|\eta^{col}_a|^2\rangle + \sum_{j \neq a,b} \langle|\eta^{col}_{a,j}|^2\rangle +
\sum_{k > j \neq a,b} \langle|\eta^{col}_{a,j,k}|^2\rangle + \ldots \bigg)
+\ \mathcal{P}_a \mathcal{A}_b \bigg( \langle|\eta^{col}_b|^2\rangle + \sum_{j \neq a,b} \langle|\eta^{col}_{b,j}|^2\rangle +
\sum_{k > j \neq a,b} \langle|\eta^{col}_{b,j,k}|^2\rangle + \ldots \bigg) \nonumber
\end{eqnarray}
\end{widetext}
where $\dots$ denote the corresponding higher order multi-body terms. 
If we prepare the qubits in a pure state, so $\mathcal{P}_a = \mathcal{P}_b=1$, the combination
\begin{equation}
\gamma^{(a)} + \gamma^{(b)} - \gamma^{(a,b)}  = {4 \over 9} \bigg( \langle|\eta^{col}_{a,b}|^2\rangle + \sum_{j \neq a,b} \langle|\eta^{col}_{a,b,j}|^2\rangle + \ldots \bigg)
\nonumber
\end{equation}
leaves only the collective coefficients involving both the qubits $a$ and $b$.

\section{Appendix C: Details on the experimental implementation}

The internal Hamiltonian of the system in the rotating frame is given by
\begin{equation}
H_{int} = \hbar \sum_{j=1}^4  \frac{\omega_{\delta,j}}{2} \sigma_z^{(j)} + \hbar \sum_{k > j =1}^4 \frac{\pi J_{j,k}}{2} \sigma_z^{(j)} \sigma_z^{(k)} 
\end{equation}
where the chemical shifts, at our 9.4 T spectrometer, are of the order of kHz:
$\omega_{\delta,1}=6650.6$ Hz, $\omega_{\delta,2}=1695.8$ Hz, $\omega_{\delta,3}=4210.0$ Hz, $\omega_{\delta,j}=-8796.7$ Hz.
The $J$-couplings are $J_{12}=72.6$ Hz, $J_{23}=69.8$ Hz and $J_{14}=7.1$ Hz, while $J_{24}=1.6$ Hz, $J_{13}=1.3$ Hz, and $J_{34}=41.6$ Hz (according 
to the characterization of the sample we used; see also \oldcite{crotonic}).\\

The experimental initial state preparation over the 4 qubits reported a correlation with the theoretical one that was on average 0.99 (0.98 the lowest). 
The correlation for the targeted qubits (a pair of qubits) was, in each case, between 1.00-0.99 (the pseudo-pure state preparation was designed to optimize 
this correlation). \\

We implemented the twirl of pairs of qubits exactly
using $36$ Clifford operators. We had to pick one of the 8 available 6-element subsets of Clifford operators for each qubit, and 
we chose the subset that performed best experimentally: we applied each candidate to the thermal equilibrium state and compared the 
experimental performance with the theoretical one. 
This criterion coincided with choosing the subset that best took the equilibrium state to the $I/2$ state
for the qubit being twirled.

To perform quantum state tomography (QST) on $4$ qubits with a liquid state NMR QIP, we used a set of $18$ readout pulses. 
Therefore the number of experiments required
to measure one collective coefficient $\langle |\eta_{a,b}^{col}|^2 \rangle$ 
for a given pair of targeted qubits $(a,b)$ for a particular gate under study was $648$, plus $18$
experiments to characterize the initial state $|00\rangle (I/2)^{\otimes 2}$ corresponding to preparing that pair in a pseudo-pure state. 
We performed QST of the full system therefore
having a broader knowledge of the experimental performance, 
but this is not required by the protocol: only the target qubits must be measured.\\

The negligibility of higher order multi-body terms in the gates under study is to be expected in liquid NMR QIP. In a simple model, 
these gates consist basically of
periods of free evolution of length 
$\tau$ (the corresponding propagator is $U_{\tau} = \exp{(- i H_{int} \tau /\hbar)}$) separated by $\pi$-pulses 
on some of the qubits (so $U_{\pi}=\exp{(- i (\pi/2+\epsilon) \sigma_{x,y}^{j})}$, already accounting for some error $\epsilon$).
For example, the sequence for the gate $I_E$ is:
\begin{eqnarray*}
\tau - \left.\pi\right)^{3,4}_x - 
\tau - \left.\pi\right)^{2}_x -
\tau - \left.\pi\right)^{3,4}_x -
\tau - \left.\pi\right)^{1,4}_x -\\
\tau - \left.\pi\right)^{3,4}_{-x} - 
\tau - \left.\pi\right)^{2}_{-x} -
\tau - \left.\pi\right)^{3,4}_{-x} -
\tau - \left.\pi\right)^{1,4}_{-x}
\end{eqnarray*}
where $\tau$ denotes free evolution for a time $\tau$, and $\left.\pi\right)^{j}_p$ denotes a $\pi$-pulse (180 degree rotation) around the
$p$-axis for the qubits $j$.  

Using the BCH formula \oldcite{bch} is straighforward to see that in the building block $U_{\tau} U_{\pi}$,
3-body and higher order terms will appear with a factor at least 
$\frac{J \tau}{\pi/2}$ smaller respect to any possible 1-body and 2-body terms. For the values of $J_{j,k}\tau$ involved in our experiment, 
we have $\frac{J\tau}{\pi/2} \leq 0.14$. This means that any possible 3-body and 4-body terms would appear with a coefficient $10$ times
smaller than the ones for 1-body and 2-body terms. 

On the other hand, the simulation of the engineered pulse sequences used in the experiment showed that all the 3-body and 4-body terms appear with 
collective coefficients $ |\eta^{col}_{j,k,j'}|^2 , \ |\eta^{col}_{j,k,j',k'}|^2 < 0.005$ for the $CNOT$ and $C_{12}(0.4)$ gates, 
and $< 0.002$ for the rest. These are much smaller than the differences
betweend measured and predicted values of $|\eta^{col}_{a,b}|^2$, which can be better explained as implementation errors in the protocol or genuine
gate errors arising from imperfect refocusing.


\begin{thebibliography}{99}
\bibitem{pauliweight} The Pauli weight (also Hamming weight) of a given $O_l$ is the number of factors in it that are not the identity $I$. 
In a quantum information processor, an operator $O_l$ with Pauli weight $w$ can cause error in up to $w$ of the $n$ qubits of the register.
\end{thebibliography}
\end{document}